\documentclass[prd,onecolumn,nofootinbib]{revtex4}
\usepackage{bm,amsmath,amssymb,graphicx}

\begin{document}

\noindent
Zhurnal Eksperimental’noi i Teoreticheskoi Fiziki {\bf 159} (3), 448 (2021)\\
Journal of Experimental and Theoretical Physics {\bf 132} (3), 374 (2021)\\

\title{Gravitational collapse of a fluid with torsion into a universe in a black hole}
\author{Nikodem Pop{\l}awski}

\altaffiliation{NPoplawski@newhaven.edu}
\affiliation{Department of Mathematics and Physics, University of New Haven, West Haven, CT, USA}

\begin{abstract}
We consider gravitational collapse of a spherically symmetric sphere of a fluid with spin and torsion into a black hole.
We use the Tolman metric and the Einstein--Cartan field equations with a relativistic spin fluid as a source.
We show that gravitational repulsion of torsion prevents a singularity and replaces it with a nonsingular bounce.
Quantum particle production during contraction helps torsion to dominate over shear.
Particle production during expansion can generate a finite period of inflation and produce enormous amounts of matter.
The resulting closed universe on the other side of the event horizon may have several bounces.
Such a universe is oscillatory, with each cycle larger in size than the previous cycle, until it reaches the cosmological size and expands indefinitely.
Our universe might have therefore originated from a black hole.
\end{abstract}
\maketitle

\noindent
{\bf Introduction}\\
The torsion tensor is the antisymmetric part of the affine connection \cite{Schr}.
The general theory of relativity (GR) assumes that this tensor vanishes \cite{GR,LL2}.
However, the conservation law for the total (orbital and spin) angular momentum of a Dirac particle in curved spacetime must be consistent with the Dirac equation that allows the spin-orbit interaction.
This consistency requires that the torsion tensor is not constrained to zero \cite{req}.
The simplest and most natural theory of gravity that extends GR by equipping spacetime with torsion is the Einstein--Cartan (EC) theory \cite{Lord,EC,non,Niko}.
In this theory, expanded by Sciama and Kibble, the Lagrangian density for the gravitational field is proportional to the Ricci scalar, as in GR.
Torsion is determined by the field equations obtained from varying the action for gravity and matter with respect to the torsion tensor \cite{Lord,EC,non,Niko}.
The torsion tensor turns out to be algebraically proportional to the spin tensor of fermionic matter, so torsion does not propagate.
Consequently, EC can be rewritten as GR with the symmetric Levi-Civita connection, in which the energy--momentum tensor of matter acquires additional terms that are quadratic in the spin tensor.
Therefore, EC is free from ghosts that can be present in other theories with torsion, in which torsion is propagating \cite{ghost}.
The multipole expansion of the conservation law for the spin tensor in EC leads to the representation of the fermionic matter as a spin fluid (ideal fluid with spin) \cite{NSH}.

Torsion can generate gravitational repulsion and prevent the formation of a cosmological singularity in a homogeneous and isotropic universe described by the Friedmann--Lema\^{i}tre--Robertson--Walker (FLRW) metric \cite{FLRW,Lord,LL2,GR} when spins of fermions are aligned, which was discovered by Hehl \cite{Hehl}, Trautman \cite{Tra}, and Kopczy\'{n}ski \cite{Kop}).
The avoidance of a singularity can occur even for randomly oriented spins because macroscopic averaging of the spin terms in the energy--momentum tensor gives a nonzero value, which was discovered by Hehl et al. \cite{HHK}.
The effective energy density and pressure of a spin fluid are given by
\begin{equation}
    \tilde{\epsilon}=\epsilon-\alpha n_\textrm{f}^2,\quad\tilde{p}=p-\alpha n_\textrm{f}^2,
    \label{intro1}
\end{equation}
where $\epsilon$ and $p$ are the thermodynamic energy density and pressure, $n_\textrm{f}$ is the number density of fermions, and $\alpha=\kappa(\hbar c)^2/32$ \cite{ApJ,HHK,NP,Gabe} with $\kappa=8\pi G/c^4$.
At lower densities, the effects of torsion can be neglected and EC effectively reduces to GR.
At extremely high densities, much greater than nuclear density, the negative corrections from the spin-torsion coupling in (\ref{intro1}) violate the strong energy condition and act as repulsive gravity that may prevent the formation of a cosmological singularity \cite{cosmo,Gabe,ApJ,iso}.
Similarly, the collapsing matter in a black hole, that can be represented by the FLRW metric, would avoid a singularity and instead reach a nonsingular bounce, after which it would expand as a new, closed universe \cite{cosmo,Gabe,ApJ,iso} whose total energy is zero \cite{energy}.

If a black hole creates a baby universe on the other side of its event horizon, then such a universe would be connected to the parent universe through an Einstein--Rosen bridge \cite{ER}.
The formation and subsequent dynamics of such a universe could not be observed outside the black hole because of the infinite redshift at the horizon.
Consequently, if our universe is closed \cite{closed}, then it might have originated as a baby universe from a bounce in the interior of a parent black hole existing in another universe \cite{Pat,ER,cosmo,ApJ}.
Quantum particle production after a bounce can generate a finite period of exponential inflation \cite{ApJ} that is consistent with the Planck observations of the cosmic microwave background radiation \cite{SD}.
A nonsingular bounce can also occur if the spin tensor is completely antisymmetric \cite{spin}.

The effects of torsion in EC are very weak and are significant in astrophysics only for black holes or the very early universe.
For example, torsion may explain the matter-antimatter asymmetry in the universe \cite{anti}.
In quantum field theory, torsion may impose a spatial extension of fermions \cite{non} (which could be tested in the future) and eliminate the ultraviolet divergence of radiative corrections represented by loop Feynman diagrams \cite{toreg}.

In this article, we consider gravitational collapse of a sphere of a homogeneous spin fluid in EC, that is initially at rest.
Such a collapse was studied in \cite{Iran}, assuming that the interval in the interior of a collapsing spin fluid is described by the FLRW metric.
That work reproduced the previous results of \cite{cosmo,iso} that showed the avoidance of singularity in spacetime represented by this metric in the presence of a spin fluid.
However, it did not explore the effects of shear that oppose torsion in preventing a singularity \cite{Kop}.
Also, the work in \cite{Iran} did not investigate what happens with a spin fluid after the bounce if an event horizon forms, focusing on the case without the horizon that is realized when the initial mass of the fluid sphere is below some threshold (this conclusion was previously reached in \cite{non}).
When an event horizon forms, the fluid cannot disperse back to the region of space outside the horizon because of the unidirectionality of the motion of matter through a horizon \cite{LL2}.
Moreover, it cannot tend to a static state because the spacetime within an event horizon is not stationary.
Consequently, the spin fluid on the other side of the event horizon must expand as a new, growing universe \cite{cosmo}.

To consider gravitational collapse of a spin-fluid sphere into a black hole, we follow a more detailed analysis of collapse of a dustlike sphere by Landau and Lifshitz \cite{LL2}, based on the work of Tolman \cite{Tol} and Oppenheimer and Snyder \cite{OS}.
This formalism relates the initial scale factor of the universe in a black hole to the initial radius and mass of the black hole.
In the absence of pressure gradients, such a collapse can be described in a system of coordinates that is both synchronous and comoving \cite{LL2}.
We use the Tolman metric \cite{Tol,OS} and the EC field equations with a relativistic spin fluid as a source.
We use the temperature to represent the energy, pressure, and fermion number density in a relativistic fluid \cite{ApJ}.
We demonstrate that, after an event horizon forms, gravitational repulsion of torsion prevents a singularity in a collapsing sphere and replaces it with a nonsingular bounce.
The resulting universe on the other side of the event horizon is closed and oscillatory with an infinite number of bounces and cycles.
Without torsion, a singularity would be reached and the metric would be described by the interior Schwarzschild solution, which is equivalent to the Kantowski--Sachs metric describing an anisotropic universe with topology $R\times S^2$ \cite{KS}.
Thanks to torsion, the universe in a black hole becomes closed with topology $S^3$ (3-sphere).

Because the presence of shear may prevent torsion from avoiding a singularity \cite{Kop}, we include quantum particle production that occurs in changing gravitational fields \cite{prod} and show that two effects appear.
During contraction, particle production with torsion act together to reverse gravitational attraction generated by shear and prevent a singularity.
During expansion, this production can generate a finite period of inflation and produce enormous amounts of matter.
Accordingly, each cycle is larger and longer then the previous cycle \cite{ApJ,ent}.
The number of bounces and cycles is finite because the universe eventually reaches a size at which the cosmological constant (which could also be explained by torsion \cite{exp}) becomes dominant and then expands indefinitely.\\

\noindent
{\bf Gravitational collapse of a homogeneous sphere}\\
For a spherically symmetric gravitational field in spacetime filled with an ideal fluid, the geometry is given by the Tolman metric \cite{LL2,Tol}:
\begin{equation}
    ds^2=e^{\nu(\tau,R)}c^2 d\tau^2-e^{\lambda(\tau,R)}dR^2-e^{\mu(\tau,R)}(d\theta^2+\mbox{sin}^2\theta\,d\phi^2),
    \label{grav1}
\end{equation}
where $\nu$, $\lambda$, and $\mu$ are functions of a time coordinate $\tau$ and a radial coordinate $R$.
We can still apply coordinate transformations $\tau\rightarrow \tau'(\tau)$ and $R\rightarrow R'(R)$ without changing the form of the metric (\ref{grav1}).
The components of the Einstein tensor corresponding to (\ref{grav1}) that do not vanish identically are \cite{LL2,Tol}:
\begin{eqnarray}
    & & G_0^0=-e^{-\lambda}\Bigl(\mu''+\frac{3\mu'^2}{4}-\frac{\mu'\lambda'}{2}\Bigr)+\frac{e^{-\nu}}{2}\Bigl(\dot{\lambda}\dot{\mu}+\frac{\dot{\mu}^2}{2}\Bigr)+e^{-\mu}, \nonumber \\
    & & G_1^1=-\frac{e^{-\lambda}}{2}\Bigl(\frac{\mu'^2}{2}+\mu'\nu'\Bigr)+e^{-\nu}\Bigl(\ddot{\mu}-\frac{\dot{\mu}\dot{\nu}}{2}+\frac{3\dot{\mu}^2}{4}\Bigr)+e^{-\mu}, \nonumber \\
    & & G_2^2=G_3^3=-\frac{e^{-\nu}}{4}(\dot{\lambda}\dot{\nu}+\dot{\mu}\dot{\nu}-\dot{\lambda}\dot{\mu}-2\ddot{\lambda}-\dot{\lambda}^2-2\ddot{\mu}-\dot{\mu}^2) \nonumber \\
    & & -\frac{e^{-\lambda}}{4}(2\nu''+\nu'^2+2\mu''+\mu'^2-\mu'\lambda'-\nu'\lambda'+\mu'\nu'), \nonumber \\
    & & G_0^1=\frac{e^{-\lambda}}{2}(2\dot{\mu}'+\dot{\mu}\mu'-\dot{\lambda}\mu'-\dot{\mu}\nu'),
    \label{grav2}
\end{eqnarray}
where a dot denotes differentiation with respect to $c\tau$ and a prime denotes differentiation with respect to $R$.

In the comoving frame of reference, the spatial components of the four-velocity $u^\mu$ vanish.
Accordingly, the nonzero components of the energy--momentum tensor for a spin fluid, $T_{\mu\nu}=(\tilde{\epsilon}+\tilde{p})u_\mu u_\nu-\tilde{p}g_{\mu\nu}$, are: $T^0_0=\tilde{\epsilon}$, $T^1_1=T^2_2=T^3_3=-\tilde{p}$.
The Einstein field equations $G^\mu_\nu=\kappa T^\mu_\nu$ in this frame of reference are:
\begin{equation}
    G_0^0=\kappa\tilde{\epsilon},\quad G_1^1=G_2^2=G_3^3=-\kappa\tilde{p},\quad G_0^1=0.
\end{equation}
The covariant conservation of the energy--momentum tensor gives
\begin{equation}
    \dot{\lambda}+2\dot{\mu}=-\frac{2\dot{\tilde{\epsilon}}}{\tilde{\epsilon}+\tilde{p}},\,\,\,\nu'=-\frac{2\tilde{p}'}{\tilde{\epsilon}+\tilde{p}},
    \label{grav4}
\end{equation}
where the constants of integration depend on the allowed transformations $\tau\rightarrow \tau'(\tau)$ and $R\rightarrow R'(R)$.

If the pressure is homogeneous (no pressure gradients), then $p'=0$ and $p=p(\tau)$.
In this case, the second equation in (\ref{grav4}) gives $\nu'=0$.
Therefore, $\nu=\nu(\tau)$ and a transformation $\tau\rightarrow \tau'(\tau)$ can bring $\nu$ to zero and $g_{00}=e^\nu$ to 1.
The system of coordinates becomes synchronous \cite{LL2}.
Defining $r(\tau,R)=e^{\mu/2}$ turns (\ref{grav1}) into
\begin{equation}
    ds^2=c^2 d\tau^2-e^{\lambda(\tau,R)}dR^2-r^2(\tau,R)(d\theta^2+\mbox{sin}^2\theta\,d\phi^2).
    \label{grav5}
\end{equation}
The Einstein equations (\ref{grav2}) reduce to
\begin{eqnarray}
& & \kappa\tilde{\epsilon}=-\frac{e^{-\lambda}}{r^2}(2rr''+r'^2-rr'\lambda')+\frac{1}{r^2}(r\dot{r}\dot{\lambda}+\dot{r}^2+1), \nonumber \\
& & -\kappa\tilde{p}=\frac{1}{r^2}(-e^{-\lambda}r'^2+2r\ddot{r}+\dot{r}^2+1), \nonumber \\
& & -2\kappa\tilde{p}=-\frac{e^{-\lambda}}{r}(2r''-r'\lambda')+\frac{\dot{r}\dot{\lambda}}{r}+\ddot{\lambda}+\frac{1}{2}\dot{\lambda}^2+\frac{2\ddot{r}}{r}, \nonumber \\
& & 2\dot{r}'-\dot{\lambda}r'=0.
\label{grav6}
\end{eqnarray}
Integrating the last equation in (\ref{grav6}) gives
\begin{equation}
    e^\lambda=\frac{r'^2}{1+f(R)},
    \label{grav7}
\end{equation}
where $f$ is a function of $R$ satisfying a condition $1+f>0$ \cite{LL2}.
Substituting (\ref{grav7}) into the second equation in (\ref{grav6}) gives $2r\ddot{r}+\dot{r}^2-f=-\kappa\tilde{p}r^2$, which is integrated to 
\begin{equation}
    \dot{r}^2=f(R)+\frac{F(R)}{r}-\frac{\kappa}{r}\int\tilde{p}r^2 dr,
    \label{grav8}
\end{equation}
where $F$ is a positive function of $R$.
Substituting (\ref{grav7}) into the third equation in (\ref{grav6}) does not give a new relation.
Substituting (\ref{grav7}) into the first equation in (\ref{grav6}) and using (\ref{grav8}) gives
\begin{equation}
    \kappa(\tilde{\epsilon}+\tilde{p})=\frac{F'(R)}{r^2 r'}.
    \label{grav9}
\end{equation}
Combining (\ref{grav8}) and (\ref{grav9}) gives
\begin{equation}
    \dot{r}^2=f(R)+\frac{\kappa}{r}\int_0^R\tilde{\epsilon}r^2 r'dR.
    \label{grav10}
\end{equation}

Every particle in a collapsing fluid sphere is represented by a radial coordinate $R$ that ranges from 0 (at the center of the sphere) to $R_0$ (at the surface of the sphere).
If the mass of the sphere is $M$, then the Schwarzschild radius $r_g=2GM/c^2$ of the black hole that forms from the sphere is equal to \cite{LL2}
\begin{equation}
    r_g=\kappa\int_0^{R_0}\tilde{\epsilon}r^2 r'dR.
    \label{grav11}
\end{equation}
Equations (\ref{grav10}) and (\ref{grav11}) give
\begin{equation}
    \dot{r}^2(\tau,R_0)=f(R_0)+\frac{r_g}{r(\tau,R_0)}.
    \label{grav12}
\end{equation}
If $r_0=r(0,R_0)$ is the initial radius of the sphere and the sphere is initially at rest, then $\dot{r}(0,R_0)=0$.
Consequently, (\ref{grav12}) determines the value of $R_0$:
\begin{equation}
    f(R_0)=-\frac{r_g}{r_0}.
    \label{grav13}
\end{equation}

\noindent
{\bf Spinless dustlike sphere}\\
Before considering gravitational collapse of a sphere composed of a spin fluid, it is instructive to consider spinless dust, for which the pressure vanishes and thus $\tilde{p}=0$.
Substituting (\ref{grav9}) into (\ref{grav11}) gives
\begin{equation}
    r_g=F(R_0)-F(0)=F(R_0),
    \label{dust1}
\end{equation}
which determines the value of $R_0$.
If $f<0$, then (\ref{grav8}) has a solution
\begin{equation}
    r=-\frac{F}{2f}(1+\cos\eta),\quad \tau-\tau_0(R)=\frac{F}{2(-f)^{3/2}}(\eta+\sin\eta),
\end{equation}
where $\eta$ is a parameter and $\tau_0(R)$ is a function of $R$ \cite{LL2,Tol}.
Choosing
\begin{equation}
    f(R)=-\sin^2 R,\quad F(R)=a_0\sin^3 R,\quad \tau_0(R)=\mbox{const.}
    \label{dust3}
\end{equation}
gives
\begin{equation}
    r=\frac{a_0}{2}\sin R(1+\cos\eta)\,\quad \tau-\tau_0=\frac{a_0}{2}(\eta+\sin\eta),
    \label{dust4}
\end{equation}
where $a_0$ is a constant \cite{LL2}.
Initially, at $\tau=\tau_0$ and $\eta=0$, the sphere is at rest: $\dot{r}=0$.
Clearly, a singularity $r=0$ is reached for all particles in a finite time.
The values of $a_0$ and $R_0$ can be determined from (\ref{grav13}), (\ref{dust1}), and (\ref{dust3}):
\begin{equation}
    \sin R_0=\Bigl(\frac{r_g}{r_0}\Bigr)^{1/2},\quad a_0=\Bigl(\frac{r_0^3}{r_g}\Bigr)^{1/2}.
    \label{dust5}
\end{equation}
An event horizon for the entire sphere forms when $r(\tau,R_0)=r_g$, that is, at $\cos(\eta/2)=\sin R_0$.

Substituting (\ref{dust3}) and (\ref{dust4}) into (\ref{grav7}) gives $e^{\lambda(\tau,R)}=a_0^2(1+\cos\eta)^2/4$.
If we define
\begin{equation}
    a(\tau)=\frac{a_0}{2}(1+\cos\eta),
    \label{dust6}
\end{equation}
then the square of an infinitesimal interval in the interior of a collapsing dust (\ref{grav5}) turns into \cite{LL2}
\begin{equation}
    ds^2=c^2 d\tau^2-a^2(\tau)dR^2-a^2(\tau)\sin^2 R(d\theta^2+\mbox{sin}^2\theta\,d\phi^2).
    \label{dust7}
\end{equation}
The initial value of $a$ is equal to $a_0$.
This metric has a form of the closed FLRW metric and describes a part of a closed universe with $0\le R \le R_0$.\\

\noindent
{\bf Spin-fluid sphere}\\
We now proceed to the main part of the article and consider gravitational collapse of a sphere composed of a spin fluid to demonstrate the formation of a nonsingular universe.
Substituting $r=e^{\mu/2}$ and (\ref{grav7}) into the first equation in (\ref{grav4}) gives
\begin{equation}
    \frac{d}{d\tau}(\tilde{\epsilon}r^2 r')+\tilde{p}\frac{d}{d\tau}(r^2 r')=0,
    \label{spin1}
\end{equation}
which has a form of the first law of thermodynamics for the energy density and pressure (\ref{intro1}) \cite{ApJ}.
If we assume that the spin fluid is composed by an ultrarelativistic matter in kinetic equilibrium, then $\epsilon=h_\star T^4$, $p=\epsilon/3$, and $n_\textrm{f}=h_{n\textrm{f}}T^3$, where $T$ is the temperature of the fluid, $h_\star=(\pi^2/30)(g_\textrm{b}+(7/8)g_\textrm{f})k_\textrm{B}^4/(\hbar c)^3$, and $h_{n\textrm{f}}=(\zeta(3)/\pi^2)(3/4)g_\textrm{f}k_\textrm{B}^3/(\hbar c)^3$ \cite{ApJ,Gabe}.
For standard-model particles, $g_\textrm{b}=29$ and $g_\textrm{f}=90$.
Since $p'=0$, the temperature does not depend on $R$: $T=T(\tau)$.
Substituting these relations into (\ref{spin1}) gives
\begin{equation}
    r^2 r'T^3=g(R),
    \label{spin2}
\end{equation}
where $g$ is a function of $R$.
Putting this equation into (\ref{grav10}) gives
\begin{equation}
    \dot{r}^2=f(R)+\frac{\kappa}{r}(h_\star T^4-\alpha h_{n\textrm{f}}^2 T^6)\int_0^R r^2 r'dR.
    \label{spin3}
\end{equation}
Equations (\ref{spin2}) and (\ref{spin3}) give the function $r(\tau,R)$, which with (\ref{grav7}) gives $\lambda(\tau,R)$.
The integration of (\ref{spin3}) also contains the initial value $\tau_0(R)$.
The metric (\ref{grav5}) depends thus on three arbitrary functions: $f(R)$, $g(R)$, and $\tau_0(R)$.

We seek a solution of (\ref{spin2}) and (\ref{spin3}) as
\begin{equation}
    f(R)=-\sin^2 R,\quad r(\tau,R)=a(\tau)\sin R,
    \label{spin4}
\end{equation}
where $a(\tau)$ is a nonnegative function of $\tau$.
This choice is analogous to a dust sphere: the first equation in (\ref{dust3}), the first equation in (\ref{dust4}), and (\ref{dust6}).
Accordingly, (\ref{spin2}) gives
\begin{equation}
    a^3 T^3\sin^2 R\cos R=g(R),
\end{equation}
in which separation of the variables $\tau$ and $R$ leads to
\begin{equation}
    g(R)=\mbox{const}\cdot \sin^2 R\cos R,\quad a^3 T^3=\mbox{const}.
\end{equation}
Consequently, we find
\begin{equation}
    aT=a_0 T_0,\quad \frac{\dot{T}}{T}+\frac{H}{c}=0,
    \label{spin7}
\end{equation}
where $a_0=a(0)$, $T_0=T(0)$, and $H=c\dot{a}/a$ is the Hubble parameter.
Substituting (\ref{spin4}) into (\ref{spin3}) gives
\begin{equation}
    \dot{a}^2+1=\frac{\kappa}{3}(h_\star T^4-\alpha h_{n\textrm{f}}^2 T^6)a^2.
    \label{spin8}
\end{equation}
Using (\ref{spin7}) in (\ref{spin8}) yields
\begin{equation}
    \dot{a}^2=-1+\frac{\kappa}{3}\Bigl(\frac{h_\star T^4_0 a^4_0}{a^2}-\frac{\alpha h_{n\textrm{f}}^2 T_0^6 a^6_0}{a^4}\Bigr).
    \label{spin9}
\end{equation}
Substituting (\ref{spin4}) into (\ref{grav7}) gives $e^{\lambda(\tau,R)}=a^2$.
Consequently, the square of an infinitesimal interval in the interior of a collapsing spin fluid (\ref{grav5}) is also given by (\ref{dust7}).

The values of $a_0$ and $R_0$ can be determined from (\ref{grav13}) and (\ref{spin4}), giving (\ref{dust5}).
Substituting them and $\dot{a}(0)=0$ into (\ref{spin8}), in which the second term on the right-hand side is negligible, gives $Mc^2=(4\pi/3)r^3_0 h_\star T^4_0$.
This relation indicates the equivalence of mass and energy of a fluid sphere with radius $r_0$ and determines $T_0$.
An event horizon for the entire sphere forms when $r(\tau,R_0)=r_g$, which is equivalent to $a=(r_g r_0)^{1/2}$.
Equation (\ref{spin9}) has two turning points, $\dot{a}=0$, if \cite{Gabe}
\begin{equation}
    \frac{r^3_0}{r_g}>\frac{3\pi G\hbar^4 h_{n\textrm{f}}^4}{8h_\star^3}\sim l_\textrm{Planck}^2,
\end{equation}
which is satisfied for astrophysical systems that form black holes.\\

\noindent
{\bf Avoidance of singularity}\\
Equation (\ref{spin9}) can be solved analytically in terms of an elliptic integral of the second kind \cite{Gabe}, giving the function $a(\tau)$ and then $r(\tau,R)=a(\tau)\sin R$.
The value of $a$ never reaches zero because as $a$ decreases, the right-hand side of (\ref{spin9}) becomes negative, contradicting the left-hand side.
The change of the sign occurs when $a<(r_g r_0)^{1/2}$, that is, after the event horizon forms.
Consequently, all particles with $R>0$ fall within the event horizon but never reach $r=0$ (the only particle at the center is the particle that is initially at the center, with $R=0$).
A singularity is therefore avoided.
Nonzero values of $a$ in (\ref{dust7}) give finite values of $T$ and therefore finite values of $\epsilon$, $p$, and $n_\textrm{f}$.

The resulting universe on the other side of the event horizon has a closed geometry (constant positive curvature).
The quantity $a(\tau)$ is the scale factor of this universe.
The universe is oscillatory: the value of $a$ oscillates between the two turning points.
The value of $R_0$ does not change.
A turning point at which $\ddot{a}>0$ is a bounce, and a turning point at which $\ddot{a}<0$ is a crunch.
The universe has therefore an infinite number of bounces and crunches, and each cycle is alike.

The Raychaudhuri equation for a congruence of geodesics without four-acceleration and rotation is $d\theta/ds=-\theta^2/3-2\sigma^2-R_{\mu\nu}u^\mu u^\nu$, 
where $\theta$ is the expansion scalar, $\sigma^2$ is the shear scalar, and $R_{\mu\nu}$ is the Ricci tensor \cite{Niko}.
For a spin fluid, the last term in this equation is equal to $-\kappa(\tilde{\epsilon}+3\tilde{p})/2$.
Consequently, the necessary and sufficient condition for avoiding a singularity in a black hole is $-\kappa(\tilde{\epsilon}+3\tilde{p})/2>2\sigma^2$.
For a relativistic spin fluid, $p=\epsilon/3$, this condition is equivalent to
\begin{equation}
    2\kappa\alpha n_\textrm{f}^2>2\sigma^2+\kappa\epsilon.
    \label{avoid1}
\end{equation}
Without torsion, the left-hand side of (\ref{avoid1}) would be absent and this inequality could not be satisfied, resulting in a singularity.
Torsion therefore provides a necessary condition for preventing a singularity.
In the absence of shear, this condition is also sufficient.

The presence of shear opposes the effects of torsion.
The shear scalar $\sigma^2$ grows with decreasing $a$ like $\sim a^{-6}$, which is the same power law as that for $n_\textrm{f}^2$.
Therefore, if the initial shear term dominates over the initial torsion term in (\ref{avoid1}), then it will dominate at later times during contraction and a singularity will form.
To avoid a singularity if the shear is present, $n_\textrm{f}^2$ must grow faster than $\sim a^{-6}$.
Consequently, fermions must be produced in a black hole during contraction.\\

\noindent
{\bf Particle production}\\
The production rate of particles in a contracting or expanding universe \cite{prod} can be phenomenologically given by
\begin{equation}
    \frac{1}{c\sqrt{-g}}\frac{d(\sqrt{-g}n_\textrm{f})}{dt}=\frac{\beta H^4}{c^4},
    \label{part1}
\end{equation}
where $g=-a^6\sin^4R\sin^2\theta$ is the determinant of the metric tensor in (\ref{dust7}) and $\beta$ is a nondimensional production rate \cite{ApJ}.
With particle production, the second equation in (\ref{spin7}) turns into
\begin{equation}
    \frac{\dot{T}}{T}=\frac{H}{c}\Bigl(\frac{\beta H^3}{3c^3 h_{n\textrm{f}}T^3}-1\Bigr).
    \label{part2}
\end{equation}
Particle production changes the power law $n_\textrm{f}(a)$:
\begin{equation}
    n_\textrm{f}\sim a^{-(3+\delta)},
\end{equation}
where $\delta$ varies with $\tau$.
Putting this relation into (\ref{part1}) gives
\begin{equation}
    \delta\sim -a^\delta\dot{a}^3.
\end{equation}

During contraction, $\dot{a}<0$ and thus $\delta>0$.
The term $n_\textrm{f}^2\sim a^{-6-2\delta}$ grows faster than $\sigma^2\sim a^{-6}$ and a singularity is avoided.
Particle production and torsion act together to reverse the effects of shear, generating a nonsingular bounce.
The dynamics of the nonsingular, relativistic universe in a black hole is described by equations (\ref{spin8}) and (\ref{part2}),
with the initial conditions $a(0)=(r_0^3/r_g)^{1/2}$ and $\dot{a}(0)=0$, that give the functions $a(\tau)$ and $T(\tau)$.
The shear would enter the right-hand side of (\ref{spin8}) as an additional positive term that is proportional to $a^{-4}$.
When the universe becomes nonrelativistic, the term $h_\star T^4$ in (\ref{spin8}) changes into a positive term that is proportional to $a^{-1}$.
The cosmological constant enters (\ref{spin8}) as a positive term that is proportional to $a^{2}$.

Particle production increases the maximum size of the scale factor that is reached at a crunch.
Consequently, the new cycle is larger and lasts longer then the previous cycle.
According to (\ref{dust5}), $R_0$ is given by
\begin{equation}
    \sin^3 R_0=\frac{r_g}{a(0)},
\end{equation}
where $a(0)$ is the initial scale factor that is equal to the maximum scale factor in the first cycle.
Since the maximum scale factor in the next cycle is larger, the value of $\sin R_0$ decreases.
As cycles proceed, $R_0$ approaches $\pi$.\\

\noindent
{\bf Inflation and end of oscillations}\\
During contraction, $H$ is negative and the temperature $T$ increases.
During expansion, if $\beta$ is too big, then the right-hand side of (\ref{part2}) could become positive.
In this case, the temperature would grow with increasing $a$, which would lead to eternal inflation \cite{ApJ}.
Consequently, there is an upper limit to the production rate: the maximum of the function $(\beta H^3)/(3c^3 h_{n\textrm{f}}T^3)$ must be lesser than 1.

If $(\beta H^3)/(3c^3 h_{n\textrm{f}}T^3)$ in (\ref{part2}) increases after a bounce to a value that is slightly lesser than 1, then $T$ would become approximately constant.
Accordingly, $H$ would be also nearly constant and the scale factor $a$ would grow exponentially, generating inflation.
Since the energy density would be also nearly constant, the universe would produce enormous amounts of matter and entropy.
Such an expansion would last until the right-hand side of (\ref{part2}) drops below 1.
Consequently, inflation would last a finite period of time.
After this period, the effects of torsion weaken and the universe smoothly enters the radiation-dominated expansion, followed by the matter-dominated expansion.

If the universe during expansion does not reach a critical size at which the cosmological constant is significant, then it recollapses to another bounce and starts a new oscillation cycle \cite{cc}.
The new cycle is larger and longer then the previous cycle \cite{ApJ,ent}.
After a finite series of cycles, the universe reaches the critical size which prevents the next contraction and and enters the cosmological-constant-dominated expansion, during which it expands indefinitely.
The value of $R_0$ asymptotically tends to $\pi$, which is the maximum value of $R$ in a closed isotropic universe given by (\ref{dust7}).
The last bounce, referred to as the big bounce, is the big bang.\\

\noindent
{\bf Final remarks}\\
If our universe is closed, then it might have been born as a baby universe in a parent black hole existing in another universe.
This hypothesis is supported by the presented analysis of gravitational collapse of a spin fluid with torsion and particle production.
A more realistic scenario of gravitational collapse should involve a fluid sphere that is inhomogeneous and rotating.
If the pressure in the sphere is not homogeneous, then the system of coordinates cannot be comoving and synchronous \cite{LL2,Lif}.
Consequently, $\nu$ and the temperature would depend on $R$ and the equations of the collapse and the subsequent dynamics of the universe would be more complicated.
If the sphere were rotating, then further complications would appear \cite{Dor} and the angular momentum of the forming Kerr black hole would be another parameter in addition to the mass \cite{Kerr}.
Nevertheless, the general character of the effects of torsion and particle production in avoiding a singularity and generating a bounce in a black hole would still be valid.\\ \\
I am grateful to my parents Bo\.{z}enna Pop{\l}awska and Janusz Pop{\l}awski for inspiring this research.\\ \\
This work was funded by the University Research
Scholar program at the University of New Haven.

\end{document}